# Calculation of the Optimal Installation Angle for Seasonal Adjusting of PV Panels Based on Solar Radiation Prediction

Yashar Naeimi[1], Farzan Kooben[2], and Mohamad Hanif Moallem[3]
[1] Isfahan University of Technology, yashar.naeimi@gmail.com
[2] Isfahan University of Technology, f.kooben@gmail.com
[3] Isfahan University of Technology, hanifmoallem@gmail.com

***Abstract:*** *An important parameter that affects PV panel performance of a solar power system is the incident solar radiation with the panel or panel's area of exposure to the sun. The direction and tilt angle of a PV panel are two important factors in PV system design. This paper itself presents the calculation of the optimum installation angles for the seasonal adjusting of solar cell panels considering apparent motion of the sun. Daily adjusting for optimal tilt angle can increase the received energy from the sun. However, making trade-off is required because changing the panel angle daily costs lots of endeavour, so we designed seasonal angles for installation of panels.*

## 1. Introduction

With the shortage of fossil fuels in our planet and potential dangers of nuclear energy, the application of renewable energies has become very popular today. They are also used in energy harvesting applications [1]. Solar energy is the most promising one between renewable energies like geothermal, or wind energy in as mochas its accessibility and abundance. Solar Energy is a free and clean energy which can be used wherever in the world by means of a photovoltaic cell [2]. Power generation using PV panels and solar water heating using collectors are commonly used nowadays. Power converters also play significant role in solar power generation, [3][4][5][6]. In both cases best performance highly depends on optimal design of collectors' orientation and tilt angle.

A study performed by Morcos showed that daily changing of panels' azimuth and tilt angles to the optimum values in Egypt achieved an annual gain in total solar radiation of 29.2% more than a fixed panel with a tilt angle equal to its geographic latitude (Mitchell and Braun, in 1983; Chen et al., 2005; Morcos, 1994; Huang and Sun, 2007; Hepbasli and Gunerhan, in 2007) [7]. Solar panels usually have a large area and are quite heavy that would cause energy loss when using a motor in a solar tracker system. Furthermore, daily and monthly adjustment of tilt angles is not a practical solution due to frequent adjustment and complex structure of required frames. Seasonal adjustment of tilt angle possibly is the simplest method practically. It is not only a practical solution but also can cause a noticeable enhancement in annual power gain of the system. In this method, the panel's tilt angle is changed 4 times in a year (seasonally) being the angle of site latitude ($\phi$) around the equinoxes and $\phi \pm \Delta$ around the solstices ($\Delta$ is the value of the tilt angle changing from the site latitude). For this method, predicting the sun's movement in the sky is an important requirement. Prediction of sun position can be achieved by investigation of the sun's apparent movement. There are many different ways for calculating the sun's apparent motion.

In this work, a simple mathematical and astronomical approach has been employed to estimate optimum panel installation angle for solar panels in each month of the year at northern hemisphere which can subsequently result optimum seasonal $\beta$ angles to be adjusted.

## 2. Methodology

The power that is received by a PV panel depends not only on the existing power in the sunlight, but also on the angle between the sun and the panel. For the optimal operation of a PV system, an important task is to install panels in the right direction and with the best tilt angle to receive maximum incident solar radiation. The module is assumed to be orienting to the equator so it faces south in the northern hemisphere and north in the southern hemisphere. To gain the maximum sunlight, the solar panel should be positioned at the angle that the solar rays of the sun arrive at the panel surface perpendicularly. If the sun's rays are perpendicular to the panel surface, the power density on the absorbing surface will be equal to the incident power density. However, when the angle between a fixed panel and the sun is continually changing, the power density on a fixed PV panel is less than that of the incident sunlight. The angle between the sun and a fixed location on Earth is related to the longitude of the site, the time of the day and year. The panel angle can be adjusted manually against the horizon seasonally. This is a relatively simple way of increasing output and does not add significantly to the cost. For mid-latitude locations, adjustment to the tilt angles every three months increases the annual energy production by about 5% [8]. The desirable tilt angle for adjusting the panels seasonally is obtained with examination of sun's position.

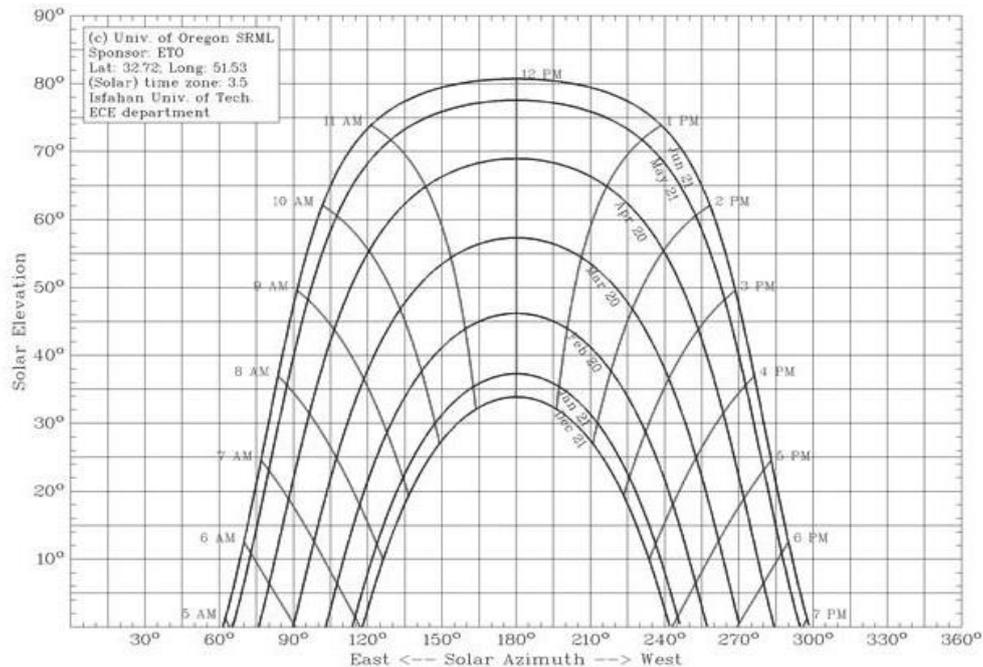

Fig. 1: Sun path charts in Cartesian coordinates at ECE department of Isfahan Univ. of Tech. location

### 3. Prediction of the Sun's Apparent Motion

The apparent motion of the sun which is due to the earth's rotation about its axis around the sun, changes the angle at which the direct component of light incidents with the Earth. If you stand on a specific location on Earth, the sun appears to move throughout the sky. This apparent motion is occurs both during a complete day and a period of 6 months. The daily movement of sun begin from the sun rise from the east to the west in northern hemisphere, but the motion of sun during a year is variation of maximum elevation angle of sun at solar noon, $\alpha$, and can be defined as the motion of sun position at solar noon in different days of a year. The important component of sun apparent motion for calculation of panels' tilt angle is the yearly one. The apparent motion of sun has been plotted by solar path chart program of Univ. of Oregon SMRL as below. In this figure we can see that the maximum elevation angle at solar noon, $\alpha$, varies between about 33° and 81° from 21 Dec to 21 Jun.

The position of the sun in the sky depends on the location of each point on Earth, the time of year, and the time of day. The Earth is tilted on its axis of rotation around the sun, so that the angle of the sun rays incident on the Earth, declination angle ($\delta$), is not zero, and varies seasonally. However, the Earth is tilted by the amount of 23.45° and the declination angle changes plus or minus this amount, and only at the fall and spring equinoxes is the declination angle equal to 0°. The amount of declination angle can be obtained by the equation [9]:

$$\delta = 23.45° \sin\left[\frac{360}{365}(d-81)\right] \quad (1)$$

Where d is the day of the year, when Jan 1st is d=1. The declination angle is zero at the equinoxes (September 22 and March 22), negative in the northern hemisphere winter and positive in the northern hemisphere summer. The declination is at its minimum of -23.45° on December 22, and maximum of 23.45° on June 22. The variation of declination angle during the different seasons cause the variation of elevation angle of sun at solar noon, maximum elevation angle, in different time of year. The elevation angle can be defined as the angular height of the sun which is measured from the horizontal.

The zenith angle, $\zeta$, at solar noon is defined as the angle between the incident sunlight at the particular location and is given by the equation:

$$\zeta = \phi - \delta \quad (2)$$

where $\phi$ is the latitude of the desired location.

### 4. Calculation of Installation Angle of Panels

The maximum elevation angle at solar noon, $\alpha$, is defined from the horizontal plane and is given by the equation as a function of latitude ($\phi$) and the declination angle ($\delta$) in northern hemisphere:

$$\alpha = 90° - (\phi - \delta) \quad (3)$$

and for the southern hemisphere:

$$\alpha = 90° + (\phi - \delta) \quad (4)$$

In equation (3) which is for the northern hemisphere, $\phi$ is positive for northern hemisphere locations and negative for locations in southern hemisphere.
The zenith angle is defined as the angle between vertical and the sun.

$$zenith = 90° - elevation \quad (5)$$

Another angle which is related to positioning of sun is Azimuth angle. The azimuth angle is the compass direction that the sunlight is coming from. In the northern hemisphere, the sun is always directly south at solar noon. The azimuth angle changes throughout the day. The elevation angle at solar noon and azimuth angle are two important factors that are used to orient PV modules. In general however, the azimuth angle varies with the latitude and time of year, and is being used to calculate the sun's position throughout the day, but it is always equal to 0° at solar noon when we calculate $\alpha$. So that for our calculations this angle is always taken as zero. In conclusion one of the two key factors for PV panel orientation, azimuth angle, is calculated equal to zero which mean the panels must be installed directly toward the south in the northern hemisphere.

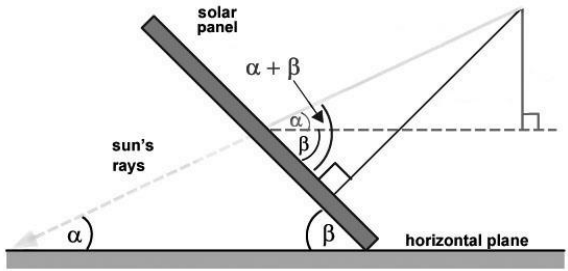

Fig. 2: Tilting the panel reduces the incoming light

As it is mentioned earlier in the article declination angle changes from 23.45 in the summer to -23.45 in the winter, and using equation (3), maximum elevation angle at solar noon can be calculated in every day throughout a year. By neglecting $\frac{360}{365}$ factor in equation (1), following equation can be obtained:

$$\delta = 23.45° \sin[(d - 81)] \quad (6)$$

From equation (3) and (6) the maximum elevation angle at solar noon can be calculated using following equation:

$$\alpha = 90° - (\phi - 23.45° \sin[(d - 81)]) \quad (7)$$

This equation gives the angle between incident photons and the Earth, $\alpha$. As can be seen in Fig. 2.
For calculating the angle of panels against the Earth, we have to attain $\beta$, where $\beta$ is the tilt angle of the panel that can be measured from the horizon. A module lying on the ground horizontally has $\beta = 0°$ and a module standing vertically has a $\beta = 90°$. $\beta$ can thus be calculated using:

$$\beta = 90° - \alpha \quad (8)$$

and;

$$\beta = \phi - 23.45° \sin[(d - 81)] \quad (9)$$

For a PV panel with a fixed tilt angle, the maximum power throughout a year is obtained when the tilt angle is equal to the latitude of the site location. However, steeper and lower tilt angle in winter and summer respectively cause a greater fraction of incident light. From equation (9) using MATLAB the variation of tilt angle in 365 consecutive days and for a specific location (in this paper our calculation is for Isfahan University of Technology with the latitude of $\phi = 32.7°$) with specific latitude can be plotted as below:

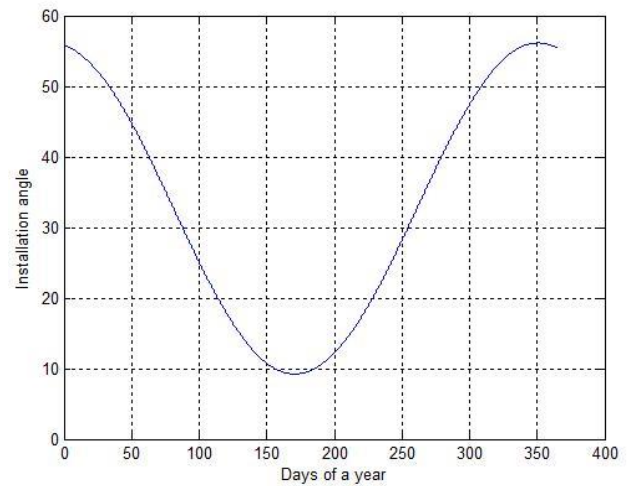

Fig. 3: Tilt angle calculated by equation (9) for 365 consecutive days at latitude of 32.7°

According to the plot, the tilt angle changes can be approximately considered to be linear. Assuming the variations as linear, we have done following computations to calculate the tilt angle for different locations on northern hemisphere and in any time of the year:
When $\delta = +23.45°$, $\alpha$ is in its maximum value as below:

$$\alpha_{max} = 90° - (\phi - 23.45°) = 113.45° - \phi \quad (10)$$

and,

$$\beta_{min} = 90° - \alpha_{max} = 90° - (113.45° - \phi) = \phi - 23.45°$$
(11)

when $\delta = -23.45°$;

$$\alpha_{min} = 90° - (\phi + 23.45°) = 66.55° - \phi \quad (12)$$

and,

$$\beta_{max} = 90° - \alpha_{min} = 90° - (66.55° - \phi) = 23.45° + \phi \quad (13)$$

As it is mentioned earlier the declination angle has a variation between +23.45° and -23.45°, considering this fact and the equation (3), *α* has a variation between 81.75° and 33.85°, as a result minimum and maximum of tilt angles of panels are 8.25° and 56.15° respectively for latitude 32.7°.

Maximum and minimum of tilt angle that were obtained in equations (11) and (13) occur in the first day of winter and summer for northern hemisphere respectively. For the days between these two we can obtain new formulas by assuming the tilt angle variation as a linear function instead of a sinusoidal one and as for the maximum and minimum values of *β*;

For January to June we have following linear equation:

$$\beta(n) = 56.15° - (n-1) \times \frac{56.15° - 8.25°}{6} \quad (14)$$

and for July to December:

$$\beta(n) = 8.25° + (n-7) \times \frac{56.15° - 8.25°}{6} \quad (15)$$

where n is the number of months with January n = 1, and *β(n)* is the tilt angle for the month n. Using equations (14), and (15), by substituting number of months, a table of tilt angles can be obtained as below:

TABLE I: Tilt angles for the months of a year

| Month | Tilt angle, *β(n)* |
|---|---|
| January | $\phi$ + 23.45° |
| February | $\phi$ + 15.47° |
| March | $\phi$ + 7.49° |
| April | $\phi$ - 0.5° |
| May | $\phi$ - 8.48° |
| June | $\phi$ - 16.46° |
| July | $\phi$ - 24.45° |
| August | $\phi$ - 16.46° |
| September | $\phi$ - 8.48° |
| October | $\phi$ - 0.5° |
| November | $\phi$ + 7.49° |
| December | $\phi$ + 15.47° |

These angles could be alternatively calculated using equation (9), or from Fig. 1. The results would be almost the same. This fact can be seen by the symmetry of calculated angles which is due to the fact that tilt angle variation is a sinusoidal function and sinusoidal functions are periodic with symmetry.

We can also obtain seasonal tilt angles from the monthly amounts of tilt angles using TABLE I for latitude of 32.7° (Isfahan University of Technology) that has been calculated as below:

TABLE II: Tilt angles for the seasonal adjustment of panels at the latitude of 32.7°

| season | Tilt angle, *β(s)* |
|---|---|
| Winter | 48° |
| Spring | 24° |
| Summer | 16° |
| Fall | 40° |

## 5. Conclusion

In this study, we discussed a straightforward approach to calculate an optimum installation angle of solar panels for locations in the northern hemisphere and for a seasonal adjustment program of panels' tilt angle. We first predict the sun's position in different times of a year by considering the sun apparent motion. Then, the different angles dealing with sun's position and panel installation including elevation angle and tilt angle were calculated. Finally the optimum tilt angle for installation and seasonal adjustment of solar panels obtained.